# Observational Constraints on the Degenerate Mass-Radius Relation

J. B. Holberg[1]

T. D. Oswalt[2]

M. A. Barstow[3]

**ABSTRACT**

The white dwarf mass-radius relationship is fundamental to modern astrophysics. It is central to routine estimation of DA white dwarf masses derived from spectroscopic temperatures and gravities. It is also the basis for observational determinations of the white dwarf initial-final mass relation. Nevertheless, definitive and detailed observational confirmations of the mass-radius relation remain elusive due to a lack of sufficiently accurate white dwarf masses and radii. Current best estimates of masses and radii allow only broad conclusions about the expected inverse relation between masses and radii in degenerate stars. In this paper we examine a restricted set of twelve DA white dwarf binary systems for which accurate (1) trigonometric parallaxes, (2) spectroscopic effective temperatures and gravities, and (3) gravitational redshifts are available. We consider these three independent constraints on mass and radius in comparison with an appropriate evolved mass-radius relation for each star. For the best determined systems it is found that the DA white dwarfs conform to evolved theoretical mass-radius relations at the 1-σ to 2-σ level. For the white dwarf 40 Eri B (WD0413-077) we find strong evidence for the existence of a 'thin' hydrogen envelope. For other stars improved parallaxes will be necessary before meaningful comparisons are possible. For several systems current parallaxes approach the precision required for the state-of-the-art mass and radius determinations that will be obtained routinely from the *GAIA* mission. It is demonstrated here how these anticipated results can be used to firmly constrain details of theoretical mass-radius determinations.

[1] Lunar and Planetary Laboratory, 1541 E. University Blvd, Sonett Space Sciences Building, University of Arizona, Tucson, AZ 85721, USA, holberg@argus.lpl.arizona.edu

[2] Florida Institute of Technology, Melbourne, FL, 32901, USA, toswalt@fit.edu

[3] Department of Physics and Astronomy, University of Leicester, University Road, Leicester, UK LE1 7RH, mab@le.ac.uk

# 1. INTRODUCTION

The theoretical mass-radius relation (MRR) for degenerate stars is widely used throughout stellar astrophysics. It allows the routine determination of white dwarf masses from the basic spectroscopic measurements of effective temperature and surface gravity, which yield such basic information as the white dwarf mass distribution. The MRR can also be used to determine the white dwarf radii necessary for the estimation of luminosities and distances, which in turn permit determination of the white dwarf luminosity function. The latter constrains the history of the Galaxy's stellar evolution and age. The MRR is also critical to both the theoretical and empirical initial-final mass relation, which describes the relationship between stellar progenitor masses and white dwarf masses and which provides constraints on how stellar matter is returned to the interstellar medium. For these reasons and a host of others it is of some importance to empirically confirm the general as well as detailed character of the theoretical MRR.

For the products of single star evolution the theoretical MRR is commonly characterized by a degenerate carbon-oxygen core, or perhaps neon-magnesium-silicon cores for white dwarfs descended from progenitors whose masses approach the core collapse limit. Overlying these degenerate cores are envelopes of helium and hydrogen that comprise white dwarf photospheres. As white dwarf masses increase from around 0.5 $M_\odot$ to the Chandrasekhar limit at ~1.4 $M_\odot$, radii vary approximately as the inverse cube root of the mass. Complicating this picture are subordinate radii dependences resulting from core composition, thermal evolution (cooling), and the 'thickness' (or relative mass fractions) of the hydrogen and helium envelopes. The ultimate goal of observational tests of the MRR is to achieve sufficient precision to identify and quantify these subordinate effects.

Empirical confirmation of the theoretical MRR has been a prime objective of numerous studies employing individual stars as well as ensembles of stars with good mass and radius determinations. It was, however, the *Hipparcos* mission (Perryman 1997) with its homogeneous determination of parallaxes for some 118,000 stars at the 1 - 2 mas level of precision that prompted the most detailed consideration of the empirical MRR. The effective *Hipparcos* faint limit of 12 to 13 magnitudes permitted only about 20 white dwarfs to be explicitly included in the *Hipparcos* input catalog. However, the parallaxes of a similar number of white dwarfs can be inferred from the parallaxes of the brighter companions of common proper motion (CPM) pairs observed by *Hipparcos*.

Anticipating the advent of *Hipparcos* parallaxes, Schmidt (1996) gave an insightful picture of the state of observational mass-radius data prior to the 1997 release of the *Hipparcos* results. Schmidt considered the empirical MRR for several sets of white dwarfs based on three observational methods: (1) surface brightness combined with surface gravity; (2) gravitational redshift plus surface gravity; and (3) a combination of (1) and (2). He found that all three methods produced a distribution of stars in the mass-radius plane clustered near 0.6 $M_\odot$ where the sharply peaked white dwarf mass distribution locates most degenerate stars. However, there was a wide dispersion of data orthogonal to the theoretical MRRs. For a small sample of stars that were part of the *Hipparcos* input catalog Schmidt found better agreement, but no solid confirmation of the theoretical MRR. Schmidt also compared theoretical MRRs using equal likelihood contours in the mass-radius plane that incorporated realistic uncertainties in the surface brightness, surface gravities, and gravitational redshifts. We will use a more precise adaptation of this method in this paper. Schmidt also identified parallax uncertainties as the primary limitation of the observational data and expressed hope that the *Hipparcos* results would greatly diminish these observational uncertainties.

Two post-*Hipparcos* evaluations of the empirical MRR, established the actual limits of the *Hipparcos* data. Vauclair et al. (1997) considered some 20 mostly bright field white dwarfs of various spectral types. The technique used was to calculate radii from *Hipparcos* parallaxes and available photometric magnitudes and then to use these radii to calculate masses from the observed surface gravity. For 17 of the stars they found generally good agreement with theoretical MRRs. Although *Hipparcos* parallaxes significantly reduced mass and radius uncertainties compared to ground-based results, the error bars remained large enough to limit a detailed comparison with theoretical MRRs. Additionally, as we shall see, some data points may well be in error.

Provencal et al. (1998) using some of the same field white dwarfs as Vauclair et al., also considered ten white dwarfs in visual binaries and common proper motion (CPM) systems where gravitational redshifts were also available in addition to surface gravities. For ten binary systems they compared masses determined by various methods including parallaxes and surface gravities, gravitational redshifts and, in two cases, dynamical orbital masses. They found relatively good agreement but also several significant outliers. The graphical comparisons with MRRs corresponding to Hamada-Salpeter zero temperature models of various core composition and several evolved models by Wood (1995) show a large scatter, with several stars apparently lying near the extreme limit defined by degenerate Fe cores. Provencal et al. found some evidence in the data for a predominance of 'thick' ($10^{-4}$ M/ $M_\odot$) hydrogen envelopes among their stars.

The above studies share some common threads. First, although *Hipparcos* parallaxes were a significant improvement over ground-based data, mass and radius error bars remained large for most stars and constituted the major source of uncertainty. Second, the number of stars was limited, with most clustered between 0.5 to 0.6 $M_\odot$, making it difficult to discern the slope of the empirical MRR. It is primarily Sirius B, with a mass near 1 $M_\odot$ and an exceptionally good parallax, which serves to define the slope of the empirical MRR. Third, although the stars have a variety of effective temperatures, the global graphical comparisons were made with respect to a limited range of zero-temperature and evolved models.

A definitive empirical confirmation of the white dwarf MRR ought to contain the following components: (1) individual mass and radii uncertainties small enough to be critically compared with theoretical expectations; (2) a sample of stars sufficiently large that statistical averages over small (0.1 $M_\odot$) mass bins can be computed; (3) a dense sampling of the mass range from 0.5 $M_\odot$ to at least 1 $M_\odot$; (4) comparisons of individual stars with theoretical MRRs of appropriate temperature. These are ambitious criteria that cannot be met by existing observations. A few well-observed stars come close to meeting the first objective. Nevertheless an existing set of suitable visual and CPM systems are potentially capable of meeting all of the above objectives. These stars will require high quality photometric, spectroscopic and gravitational redshift observations and above all, accurate trigonometric parallaxes that have relative uncertainties better than 1%. A focused program of ground-based observations will be required before accurate parallaxes from the *GAIA* mission become available a decade from now.

*GAIA* parallaxes (Garcia-Berro et al. 2005) will be anywhere from a factor of 20 to 50 times more precise than *Hipparcos* parallaxes and will extend to stars some six magnitudes fainter. When available, *GAIA* parallaxes will effectively eliminate a current major source of uncertainty in the determination of white dwarf radii. However, in order to fully benefit from *GAIA* data it will be necessary to have a set of spectroscopic and photometric data on appropriate white dwarf binary systems that minimize the other (non-parallax) sources of observational uncertainty. For several systems current data, including parallaxes, are already approaching this state (see Section 5). However, for other systems the precision of existing surface gravities, effective temperatures, multi-band photometry and gravitational redshifts must be improved. One objective of our project is to define a set of some 100 white dwarf binary systems covering a range of white dwarf masses that have the potential to provide high quality white dwarf masses and radii when *GAIA* parallaxes become available.

In this paper we make use of several enhancements to the existing data that have become available since the initial discussions of the *Hipparcos* parallaxes: (1) the revised *Hipparcos* parallaxes (van Leeuwen 2007) and in several cases statistical averages of ground-based and *Hipparcos* parallaxes; (2) new *HST* spectroscopy results for Sirius B (Barstow et al. 2005) and a recently analyzed Sirius-like system, ε Ret B, (Farihi et al. 2011); (3) the new photometric zero-points for ground-based photometry on the *HST* photometric scale (Holberg & Bergeron 2006); and (4) a set of individualized self-consistent evolved MRRs[1]. All of these additions and enhancements lead to a much more consistent and uniform treatment of the observational data for testing the theoretical MRRs.

In Section 2 we discuss our sample of stars and their existing parallax, photometric, temperature, gravity and gravitational redshift data. In Section 3 we discuss the theoretical MR relations we will test and in Section 4 the methods used to make independent estimates of white dwarf masses and radii. In Section 5 we present comparisons of our results for individual stars with theoretical MRRs. We conclude in Section 6 with discussions of the current state of the observational verification of the degenerate MRRs, future observations planned for additional systems, and methods for improving observational testing of MRRs.

---

[1] www.astro.umontreal.ca/~bergeron/CoolingModels/

## 2. DATA

The white dwarf stars considered here are members of binary systems that have been drawn from several sources. Some can be found in the local 20 pc population of white dwarfs (Holberg et al. 2008). Others are members of the wider sample of white dwarf components of visual and CPM binaries (for example, Oswalt, Hintzen, & Luyten 1988). Others were taken from lists of gravitational redshift measurements, such as Koester (1987), Reid (1996) and Silvestri et al. (2001). All are resolved systems containing DA white dwarfs that have good parallaxes, photometry, spectroscopic temperatures, gravities and gravitational redshift measurements. These systems are listed in Tables 1 and 2. In addition there exists a larger set of resolved binaries where one or more of these measurements are currently not available, or are perhaps suspect. These systems have the potential to meet our observational requirements.

We focus exclusively on DA stars since the analysis of the HI Balmer line profiles yields confident estimates of $T_{eff}$ and log g. Further, the Hα, and to a lesser extent Hβ profiles, provide reliable estimates of the gravitational redshift. This excludes such well-known systems as Procyon B, which has an historical orbit, but is known to be a DQZ star (Provencal et al. 2002.). We also currently restrict our range of $T_{eff}$ to DA stars hotter than ~12,000 K, since it is clear from the work of Bergeron, Leggett, & Ruiz (2001) and others that the spectroscopic surface gravities for cooler stars are biased towards higher gravities. The existence of such a bias is a potential confounding factor in our use of surface gravities in this paper and until the origin of this effect is better understood (Tremblay et al. 2011) we have not included cooler DA stars.

*2.1 Parallaxes*

Accurate trigonometric parallaxes are fundamental to the determination of white dwarf masses and radii. There presently are about 380 published parallax measurements for white dwarfs. Here we use two main sources of such parallaxes: The *General Catalog of Trigonometric Stellar Parallaxes* (Van Altena et al. 1994, hereafter 'Yale parallaxes') and the *Hipparcos* parallaxes (Perryman 1997). The latter source is superseded by the newer reductions of van Leeuwen (2007). The revised van Leeuwen (2007) *Hipparcos* parallaxes have significantly reduced parallax uncertainties compared to the original 1997 catalog. For some stars such as Sirius B, 40 Eri B and others we computed a weighted average of the various parallax measurements, when they were consistent. In those cases where conflicts exist between Hipparcos and Yale parallaxes we present the consequences adopting the alternate parallaxes. Also since we are dealing with binary systems the parallaxes used are often those of the bright main sequence component and not the white dwarf. Such parallaxes are likely to be more accurate than those derived from faint white dwarf components.

*2.2 Photometry*

Accurate photometry goes hand-in-hand with accurate parallaxes for precise radii determinations. Frequently only V magnitudes are available. However, for many stars UBVRI + $JHK_s$, *ugriz*, as well as Strömgren *ubyv* photometry is now available. Wherever possible in this paper we use the results of Holberg & Bergeron (2006) who obtained photometric measurements on the *HST* photometric system (Bohlin & Gilliland 2004) together with a defined set of magnitude offsets and photometric zero points. This significantly reduces radii biases that can arise from use of heterogeneous photometric data.

*2.3 Spectroscopic Temperatures and Gravities*

It is now routine to derive spectroscopic temperatures and gravities for DA white dwarfs from detailed fitting of Balmer line profiles. Dozens of large surveys now provide such data for thousands of DA stars, for example, Liebert, Bergeron & Holberg (2005). Our adopted effective temperatures and gravities (and references) are given in Table 1. For the sake of consistency regarding various sources of models, line broadening theory, and data reduction techniques we have, wherever possible, used results that employed the techniques outlined in Bergeron, Liebert, & Holberg.

*2.4 Gravitational Redshifts*

The prediction that massive bodies produce non-Doppler redshifts proportional to their mass was one of three original observational tests of Einstein's General Theory of Relativity. Indeed, when the gravitational redshift was first observed in Sirius B by Adams (1925) it was hailed as a confirmation of general relativity. This measurement, although ultimately shown to be in error, was also the first widely accepted evidence for the existence of collapsed stars – white dwarfs (Holberg 2010). Today white dwarf redshifts are frequently used to determine white dwarf radii and masses.

The advent of large aperture telescopes and efficient high dispersion spectroscopy has made the determination of white dwarf gravitational redshifts relatively easy. Determination of gravitational redshifts involves establishing a velocity reference frame that defines kinematic and orbital motion from which the gravitational redshift must be distinguished. This could, for example, be the local standard of rest, which yields a determination of an ensemble average redshift with respect to the apparent velocities of a set of field white dwarfs (see the early work of Trimble & Greenstein 1967 and Greenstein 1972 and more recently, Falcon et al. 2010). Alternatively, the reference frame could be the space motion of an open stellar cluster such as the Hyades where redshifts of individual white dwarfs can be obtained (Reid, 1996). The most frequently employed method is the use of resolved binary systems containing a white dwarf where the radial velocities of both components can be determined. This is the approach taken in this paper.

In this paper we primarily rely on four such studies. Koester (1987) measured the gravitational redshifts of nine DA stars using the echelle spectrograph on the 3.6m ESO Telescope. These were wide binary systems in which the non-LTE (NLTE) cores of the Hα lines in the white dwarfs were used, avoiding possible line asymmetries in the broader wings (Schulz 1977). Wegner & Reid (1991) presented redshift results for 23 white dwarf members of common proper motion systems. These measurements were obtained with the double spectrograph on the 200-inch Hale Telescope and employed the NLTE cores of the Hα and Hβ lines. No corrections for the gravitational redshift of the main sequence companions were applied. Reid (1996) measured the gravitational redshifts of some 53 white dwarfs in wide binaries, as well as members of the Hyades and Praesepe clusters. Reid used the HIRES spectrograph on the Keck Telescope and determined redshifts with respect to the NLTE line cores for Hα and

Hβ lines and included a uniform +0.5 km s$^{-1}$ correction for the gravitational redshift of the main sequence companion. We focus here on the 28 non-cluster systems studied by Reid. Silvestri et al. (2001) determined the gravitational redshifts for a sample of 41 white dwarfs in wide binaries using the NOAO 4m telescopes at Kitt Peak and CTIO and the HIRES spectrometer on the Keck Telescope. In most cases Hα line cores were used but no corrections for the gravitational redshift of the main sequence companions were applied.

## 3. THEORETICAL MASS-RADIUS RELATION

*3.1 Theoretical Mass-Radius Relation*

The relationship between the mass of a white dwarf and its radius, as well as the limiting mass of such a star, were ideas developed from the relativistic equation of state for a degenerate electron gas in the early 1930s by S. Chandrasekhar and E. C. Stoner (Nauenberg, 2008). Attempts to observationally confirm this relationship were initially hampered by an incorrect radius for Sirius B resulting from Adams' (1925) erroneous gravitational redshift, which was dominated by scattered light from Sirius A, and were compounded by a long running dispute between Chandrasekhar and Eddington about the relativistic and non-relativistic equations of state which yield two very different mass-radius relations (Wesemael 2009). Prior to the Second World War the closest anyone came to observationally resolving this issue was G. Kuiper in 1939 (Holberg 2009). Critical considerations of the MRR resumed in 1961 with the publication of the Hamada-Salpeter theoretical zero-temperature MRRs of Hamada & Salpeter (1961) for white dwarfs of various internal compositions. Many subsequent comparisons of observed masses and radii have used the zero-temperature Hamada-Salpeter relations.

Actual white dwarfs, however, are in a continual state of thermal evolution (cooling). Their precise radii are influenced not only by internal composition but also their declining surface temperatures and the masses of any helium and/or hydrogen envelopes that exist. It is therefore most appropriate to compare observed radii and masses with thermally evolved models of the correct temperature. The most frequently used models are those of Wood (1995). For this paper we use the MRRs implicit in the Bergeron photometric DA model grid and directly calculate the radius vs. mass relation for the defined temperature and gravity of each star from interpolations with respect to the tabulated values in the photometric grid. The Bergeron photometric grid assumes the carbon core Wood models with a 'thick' fractional hydrogen layer of $10^{-4}$ stellar masses above 30,000 K and the cooling models of Fontaine, Brassard & Bergeron (2001) below 30,000 K. The latter models also assume fractional hydrogen layer masses of $10^{-4}$ $M_*$ but with carbon-oxygen cores.

## 4. METHODS

Binary systems containing white dwarfs offer an unique opportunity to independently determine the masses and radii of degenerate stars. In fortunate circumstances such as Sirius A-B and Procyon A-B extended astrometric observations of binary orbits also permit an accurate estimate of the white dwarf masses. The corresponding radii, on the other hand, must be inferred from photometric or spectroscopic observations. The radius can viewed be as dependent variable that is primarily a function of the total stellar mass, but which is also a subordinate function of the composition of the degenerate core, surface temperature and the structure of the non-degenerate envelope.

The method we adopt is similar to that used by Schmidt (1996). The stellar radius is independently determined by three relations having different mass dependence: (1) the photometric equation linking white dwarf surface brightness with the observed flux at the top of the Earth's atmosphere; (2) the spectroscopic determination of surface gravity; and (3) the gravitational redshift. Relation (1) is virtually independent of mass, although the Eddington flux has a small dependence on gravity. Relation (2) depends on $M^{1/2}$. Relation (3) depends directly on $M$.

Beginning with the photometric relation for a DA white dwarf,

$$f_\lambda = 4\pi (R^2/D^2) H_\lambda (T_{eff}, \log g) \ , \tag{1}$$

where $R$ is the stellar radius, $D$ the trigonometric parallax distance, and $H_\lambda$ is the monochromatic Eddington flux at wavelength $\lambda$, which for a DA white dwarf depends on the surface temperature $T_{eff}$ and the surface gravity $\log g$. Over a wide range of temperatures the latter two quantities can be determined spectroscopically without reference to photometry. The Eddington flux is determined by radiative transfer in the stellar photosphere while the stellar radius is a quantity that implicitly depends on the interior physics of the white dwarf *i.e.* total mass, core composition, surface temperature and the masses of any overlying envelopes of He and H. This radius can be expressed as

$$R_1 = D\sqrt{\frac{f_\lambda}{4\pi H_\lambda(T_{eff},\log g)}}, \qquad (2)$$

and $f_\lambda = f_0 10^{-0.4 m_\lambda}$,

where $f_0$ is the photometric zero point for magnitude $m_\lambda$. For example, for V-band magnitudes and distances (in pc) the radius (in solar radii) is expressed as

$$R_1 = 771.4 D (10^{-0.2 m_v})(H_V(T_{eff},\log g))^{-1/2}. \qquad (3)$$

Since absolute monochromatic fluxes or magnitudes are seldom available it is standard practice to use broad-band photometric fluxes and magnitudes. The techniques for doing this and for determining a consistent set of photometric zero-points on the *HST* photometric scale are described in Holberg & Bergeron (2006). Alternatively, it is sometimes more precise to use a set of independent apparent stellar magnitudes at different wavelengths in the photometric relation to determine the stellar solid angle $\Omega = \pi(R^2/D^2)$ by fitting the stellar energy distribution. In this case $R_1$ can be rewritten as

$$R_1 = D\sqrt{\Omega/\pi}, \qquad (4)$$

From the spectroscopic surface gravity, $g$, the radius can be expressed as

$$R_2 = \sqrt{\frac{g_{sun}}{g} M}, \qquad (5)$$

where $g_{sun}$ is the solar surface gravity, $g_{sun} = GM_\odot/R_\odot^2 = 10^{4.438}$, in cgs units. Finally the gravitational redshift gives

$$R_3 = \frac{V_{sun} M}{V_{gr}}, \qquad (6)$$

where $V_{gr}$ is the observed gravitational redshift velocity in (km s$^{-1}$) and $V_{sun}$ is the solar gravitational redshift, $V_{sun} = GM_\odot/cR_\odot = 0.6360$ km s$^{-1}$, where $c$ is the speed of light.

These three independent radii constraints can be combined into a two-dimensional $\chi^2$ function which can be minimized with respect to mass and radius,

$$\chi^2 = \left(\frac{R-R_1}{\Delta R_1}\right)^2 + \left(\frac{R-R_2}{\Delta R_2}\right)^2 + \left(\frac{R-R_3}{\Delta R_3}\right)^2, \tag{7}$$

where the quantities $\Delta R_1$, $\Delta R_2$ and $\Delta R_3$ are the respective uncertainties in $R_1$, $R_2$ and $R_3$ computed from the uncertainties in the observed quantities $T_{eff}$, $\log g$, $m_\lambda$, $D$, and $V_{gr}$. These radius uncertainties can be expressed as

$$\Delta R_1 = R_1 \sqrt{\left(\frac{\Delta D}{D}\right)^2 + \left(0.2\ln(10)\Delta m_\lambda\right)^2 + \left(\frac{\Delta H_\lambda}{2 H_\lambda}\right)^2}, \tag{8}$$

or alternately

$$\Delta R_1 = R_1 \sqrt{\left(\frac{\Delta D}{2D}\right)^2 + \left(\frac{\Delta \Omega}{2\Omega}\right)^2}, \tag{9}$$

$$\Delta R_2 = R_2 \left(\tfrac{1}{2}\ln(10)\Delta \log g\right), \tag{10}$$

$$\Delta R_3 = R_3 \left(\frac{\Delta V_{gr}}{V_{gr}}\right), \tag{11}$$

*Alternative Methods*

For isolated white dwarfs – those not in open clusters or binary systems – it is possible to independently measure masses and radii by obtaining accurate trigonometric parallaxes and spectroscopic determinations of effective temperature and surface gravity. Basically, in the mass-radius plane it is the parallax, coupled with photometry and effective temperature that determines a radius constraint that is independent of mass. The surface gravity then provides a separate constraint on $M/R^2$. By this means a mass and radius can be determined and the corresponding uncertainties estimated. Vauclair (1997) and Provencal et al. (1998) took advantage of accurate *Hipparcos* parallaxes for approximately a dozen white dwarfs to compare the resulting masses and radii with theoretical mass-radii relations. Even with the improved van Leeuwen (2007) *Hipparcos* parallaxes, the resulting radii uncertainties were 5% – 10% and slightly greater for mass uncertainties.

There are other paths to testing the MRR. For example, both orbital dynamics and asteroseismology can be used to determine the masses of certain specific white dwarf stars but parallaxes, photometry and spectroscopy are still required to determine the corresponding radii. A current series of observations with *HST* (Bond 2009) will soon yield dynamical masses for white dwarfs in several binary and double degenerate systems containing white dwarfs. However, of these, only the Sirius A–B system contains a DA white dwarf. The radii still must be determined from their stellar surface brightness and trigonometric parallaxes. It has proved possible through the asteroseismological study of pulsating ZZ Ceti and other types of non-radial pulsating white dwarfs to estimate pulsational masses (Winget & Kepler 2008). However, in these cases pulsational masses are sensitive to the identified $\ell$ mode, the H and He envelope masses, and core composition in addition to the overall mass. For example, Pech, Vauclair & Dolez (2006) found a mass of $0.575 \pm 0.005$ $M_\odot$ for HL Tau 76 (WD0415+271). Benvenuto et al. (2002) found a mass of 0.525 $M_\odot$ for G117-B15A (WD0921+354).

Eclipsing binaries in principle can yield both white dwarf masses and radii. These systems are almost invariably close post-common envelope (CE) systems. The radii can be determined from detailed modeling of ingress and egress light curves and the masses from orbital dynamics. Examples of such systems are NN Ser (WD1550+130) and RR Cae (WD0419-487), which contain a white dwarf and a dM star with orbital periods of 3.1 hrs and 7.3 hrs, respectively. For NN Ser Parsons et al. (2010) found a mass of $0.535 \pm 0.012$ $M_\odot$ and a radius of $0.0211 \pm 0.0002$ $R_\odot$ and for RR Cae of Maxted et al. (2007) find a mass of $0.440 \pm 0.022$ $M_\odot$ and a radius of $0.0155 \pm 0.00045$ $R_\odot$. For both systems the masses and radii match well the predictions of theoretical MRRs although RR Cae is a He-core white dwarf. Even though such white dwarfs bear the imprint of post-common envelope evolution and are not representative of field white dwarfs they are potentially good sources of information on common envelope mass loss. However, even in these cases where eclipses constrain the system inclination, unless both the white dwarf transit and eclipse can be actually measured the masses and radii retain some inclination dependence.

## 5. ANALYSIS

In Table 1 we list twelve DA systems and the corresponding observational parameters. Included for each object are the standard 'WD' catalog designation (McCook & Sion 1999), an object name, $T_{eff}$, log g, V magnitude, parallax and gravitational redshift. Also included are uncertainties and references. In instances where multiple references are available weighted averages have been computed. The non-degenerate companion star name together with relevant information such as spectral type, V magnitude, angular and physical separations, and estimated orbital periods are given in Table 2. For the M star companions where published spectral types were discrepant we have reconciled our spectral types in Table 2 with the distance moduli and observed 2MASS $JHK_s$ magnitudes with the absolute J, H and $K_s$ magnitudes. Also in Table 2 the estimated periods are the minimum orbital periods computed from Kepler's Third Law and the sum of the stellar masses. In Figures 1a-l we plot minimum 1-σ and 2-σ $\chi^2$ contours in the mass-radius plane that correspond to the observational parameters and uncertainties given in Table 1. Also shown in Figs. 1a-l are the theoretical evolved MRRs (see Section 3) appropriate to the $T_{eff}$ and log g of each DA.

***WD0413-077 (Fig 1a):*** This is the well-known white dwarf 40 Eri B. It is in a hierarchical triple system (Heintz 1974) containing 40 Eridani A (HR 1325, K0.5V) and 40 Eri. C (M5Ve). The gravitational redshift and separations discussed here are with respect to 40 Eri C. In Fig.1a 40 Eri B lies ~ 2-σ below the 'thick' envelope MRR. We have also included in this plot the range of dynamical mass of 0.501 ± 0.011 $M_\odot$ from Shipman et al. (1997) as a vertical region. Shipman et al. obtained their mass from Heintz (1974) who had estimated a mass of 0.43 ± 0.02 $M_\odot$. Shipman et al. derived the larger mass using a *Hipparcos* revision of Heintz's parallax. It had been of some concern that the low Heintz mass seemed inconsistent with single star evolution. Shipman et al. resolved this paradox by noting the *Hipparcos* parallax implied a larger mass. As can be seen here our best fitting mass of 0.51 $M_\odot$ is in agreement with the dynamical mass, but the radius remains below the 'thick envelope' MRR. If we instead use the 'thin envelope' of Wood (1995), plotted in Fig. 1a as a dotted line, a much better match is achieved. Our estimated mass and radius, 0.51 $M_\odot$ and 0.0135 $R_\odot$, are in very good agreement with the results of Shipman et al. 0.501 ± 0.011 $M_\odot$ and 0.0136 ± 0.00024 $R_\odot$.

*WD0416-594 (Fig. 1b):* This is a recently identified DA white dwarf (Chauvin et al. 2006) orbiting the bright star ε Ret (K2IV, HD 27442). The main sequence component also hosts an exoplanet (Butler et al. 2006, 2001). The white dwarf component is at a separation of 13". Recently, Farihi et al. (2011) obtained uncontaminated VLT UVES spectra of this companion (designated WD0415-594). These results provide the first gravitational redshift, spectroscopic $T_{eff}$, and log g for the white dwarf, yielding a mass and radius in good agreement with theory (Farihi et al. 2011). Although we use the data from Farihi et al. in this analysis, our mass-radius plane contours differ slightly due to a more consistent computation of the uncertainty ellipses. It is evident that ε Ret B is in satisfactory agreement with the MRR at the 1-σ level. The primary contribution to the uncertainty is the assumed V magnitude, which can in principal be improved from ground-based observations in the near IR.

*WD0642-166 (Fig. 1c):* This is the prototype white dwarf Sirius B and the most massive star in our sample. Sirius B has been the subject of numerous studies over the years but detailed observations of its photometric magnitude, spectrum and gravitational redshift have been hampered by the proximity to the overwhelmingly bright Sirius A. Barstow et al. (2005) finally eliminated the longstanding problem of scattered light by observing Sirius B with the *Space Telescope Imaging Spectrograph* (*STIS*) on the *Hubble Space Telescope* (*HST*). These observations gave an accurate gravitational redshift (80.42 ± 4.83 km s$^{-1}$), an improved spectroscopic temperature and gravity ($T_{eff}$ = 25,193 ± 37 K and log g = 8.556 ± 0.01). We use here the conventional apparent magnitude of 8.44 ± 0.06 since the *HST* magnitude estimate of V = 8.528 has a highly asymmetrical error estimate biased to the bright side. Fig. 1c shows the mass-radius contours for Sirius B, along with the best existing estimate for its dynamical mass of 1.019 ± 0.013 M$_\odot$ (Schaefer, et al. 2006) derived from an on-going analysis of *HST* orbital data. As can be seen, both the dynamical mass and the best fitting mass and radius for this star are in close agreement with the theoretical MRR. *Sirius B is the only star in our sample that currently fixes the high mass end of the MRR.* It should be noted that planned *HST* observations should achieve a more accurate gravitational redshift and better photometry that have the potential to significantly reduce the existing error ellipses. In Fig. 1c note the shift in the mass and radius plot axes to cover the larger mass and smaller radius of Sirius B.

***WD1105-048 (Fig 1d):*** This is a DA3.2 star in a binary system with an M3V (LP 672-2) companion with a 279" separation. Figure 1d shows a best-fitting solution that includes the theoretical MRR at the 1-σ level. However, the significance of this largely diminished by the size of the error ellipses. This uncertainty is totally dominated by the existing Yale parallax of 22.5 ± 7.2 mas. The LP 672-1/2 system represents somewhat of a dilemma. Although the Yale parallax for the white dwarf has a 'G' (good) quality flag the Yale parallax for the companion is 57.7 ± 12.4 mas with no quality flag. Hence the existing trigonometric parallaxes are in clear conflict. Holberg et al. (2008) computed a consistent UBVRI photometric distance of 25.19 ± 0.32 pc ($\pi$ = 39.7 ± 0.5 mas) for the white dwarf. Likewise, for the M3 star we find a consistent set of absolute $JHK_s$ magnitudes for a parallax of 57.7 mas. In summary, a meaningful comparison with the MRR will only be possible when this parallax discrepancy is resolved and the uncertainty is improved.

***WD1143+321 (Fig 1e):*** This is a DA3.4 star with an M3V (*BD-18° 3019*) companion with a 9" separation. In Fig. 1e the error ellipses are marginally consistent with the MRR at the 2-σ level. The Yale parallax and the V-magnitude uncertainties dominate the existing solution.

***WD1314+293 (Fig. 1f):*** This is the well-known hot DA star *HZ43,* which is approximately 2.2" away from its active M3.5V companion *HZ43B*. We have used the Yale parallax of 15.3 ± 2.3 mas rather the *Hipparcos* parallax of 25.96 ± 6.38 because the latter does not provide an acceptable match to the MRR. This discrepancy was first noticed by Vauclair et al. (1997) but its cause remains unknown. Although the 1-σ error ellipse includes the MRR, existing data cannot be used to constrain any models.

***WD1327-083 (Fig. 1g):*** This is the white dwarf *Wolf 485A* which has a very wide M4 CPM companion (*Wolf 485B*) at an 11 arc min. separation. Our solution, 0.54 $M_\odot$ and 0.0141 $R_\odot$, sits on the MRR for this star. Indeed, considered individually, the photometric, log g, and gravitational redshift constraints cross the MRR at respective masses of 0.54 $M_\odot$, 0.53 $M_\odot$, 0.55 $M_\odot$, giving a very well constrained mass and radius for this white dwarf. However, this apparent agreement must be considered somewhat fortuitous given the size of the error ellipses. This star needs an improved parallax and gravitational redshift.

***WD1620-391 (Fig. 1h):*** This the well-known DA white dwarf *CD-38° 10980* which is part of the wide Sirius-like system containing the G5V star *CD-38° 10983* with a separation of 345". This system also contains a Jovian-mass exoplant at 1.3 AU (Butler et al. 2006). The plot in Fig. 1h shows remarkably small error ellipses sitting astride the MRR. Here we have used the well-determined gravitational redshift of Silvestri et al. (2001) of 33.9 ± 0.4 km s$^{-1}$. If we had instead used the Koester (1987) redshift of 37.9 ± 2.0 km s$^{-1}$, the resulting mass would be 0.08 M$_\odot$ greater, with much larger error ellipses that place the star over 2-$\sigma$ above the MRR. The parallax used is a weighted mean of the *Hipparcos* parallaxes for the white dwarf and its G5V companion, $\pi$ = 78.21 ± 0.37 mas. Yale parallaxes also exist for both stars but their weighted mean of $\pi$ = 64.78 ± 6.07 mas is inconsistent with the *Hipparcos* parallax. If the Yale parallax is used both the mass and radius increase and become totally inconsistent with the MRR. Likewise, Holberg et al. (2008) noted the incompatibly of the Yale parallax and the photometric distance to the DA star.

***WD1706+332 (Fig. 1i):*** This is DA3.9 white dwarf *G181-B5B* with an F8 companion (*BD+33° 2834*) at a separation of 35.6". It is one of the systems discussed by Provencal et al. (1998). Our determined mass of 0.54 M$_\odot$ and radius of 0.0125 R$_\odot$ are somewhat greater than those of Provencal et al. (0.50 ± 0.05 M$_\odot$ and radius 0.0119 ± 0.001 R$_\odot$) and is in marginally better agreement with the MRR. Our results are largely on the van Leeuwen (2007) parallax of 14.85 ± 0.81 mas for the F8 companion (the 1997 *Hipparcos* parallax employed by Provencal et al. was 14.6 ± 1.0 mas). It should also be pointed out that the $\chi^2$ statistic for this star is very large (26.4) indicating an inconsistency between the three methods of estimating the radius. It is the parallax which produces the small radius.

***WD1716+020 (Fig. 1j):*** This is the DA white dwarf (*Wolf 672A, LHS 3278*) which has an M3.5 companion (*LHS 3279*) at a separation of 13.1". Fig. 1j shows that the mass and radius is consistent with the MRR at the 1-$\sigma$ level. Here we have used the Yale parallax of 28.2 ± 2.6 mas. An alternate parallax of 19 ± 2 mas by Smart et al. (2003) is totally inconsistent with the MRR. Smart et al. argue the larger Yale parallax is the result of an improper averaging of the parallaxes of the white dwarf and the M5 companion. As with WD1706+382 the Yale solution has a very large $\chi^2$ statistic (24.4) with much of the uncertainty attributable to the parallax.

***WD1743-132 (Fig. 1k):*** This is the DA white dwarf (*G154-B5B*) which has an M1V companion (*G154-B5A*) at a separation of 32.2". Our optimal mass of 0.46 ± 0.06 $M_\odot$ and radius 0.0129 ± 0.0009 $R_\odot$ is similar to those of Provencal et al. (0.46 ± 0.08 $M_\odot$ and radius 0.0130 ± 0.002 $R_\odot$). While the star is near the MRR in Fig. 1k the error ellipses are large and do not well constrain the models.

***WD2341+322 (Fig. 1l):*** This is a wide binary where the white dwarf is *LP 347-6*. It is associated with the M1.5 CPM companion (*LP347-5*) at a separation of 174". Although this system has a *Hipparcos* parallax it was not among those systems considered by Provencal et al. (1998) or Vauclair et al. (1997). Our mass of 0.56 $M_\odot$ and radius of 0.0124 $R_\odot$ is consistent (within 1-$\sigma$) with the MRR.

Fig. 2 is a global display of the results for all twelve white dwarfs contained in Fig. 1. In Fig. 2 the theoretical MRR is defined to be unity and the observed radii and 1-$\sigma$ error bars are represented as ratios with respect to unity. The orientation of the error bars follows the major and minor axes of the 1-$\sigma$ error ellipses in Fig. 1. Table 3 summarizes the mass and radius results (in solar units) including uncertainties from our analysis including the $\chi^2$ statistic. Also included are the photometrically determined radii ($R_\pi$) together with the masses where the photometric, surface gravity and gravitational redshift constraints cross the theoretical MRRs. Effectively, these are the masses that would be measured by each of the three methods separately.

In summary, nine of the stars that we considered have observed masses, radii and associated uncertainty ellipses that are consistent at the 1-$\sigma$ level or better with temperature-appropriate evolved theoretical MRRs having 'thick' H envelopes. Except for the case of 40 Eri B, it is not possible using the present data to distinguish a clear preference for 'thick' or 'thin' models, other than to note that 'thin' models do not improve overall the results shown in Fig.1. Three of these stars (WD1105-048, WD1314+293 and WD1743-132), while formally consistent with theoretical MRRs at the 1-$\sigma$ level, offer no meaningful constraints because of large observational uncertainties. Our result for the star 40 Eri B clearly favors a 'thin' H envelope. The case for 40 Eri B actually having a 'thin' envelope is strengthened by the fact that the parallax and V magnitude are very well determined and effectively fix the radius within a very narrow range. In addition, the dynamical mass also serves to locate the star well below the 'thick' envelope

MRR. Thus, it is very difficult to place 40 Eri B near the theoretical 'thick' envelope MRR without violating either or both the photometric and orbital mass constraints. One of our stars (WD1143+321) is marginally consistent with the MRRs at the 2-σ level.

## 6. CONCLUSIONS AND FUTURE RESULTS

Our results arguably provide an improved foundation for a systematic confirmation of contemporary evolved white dwarf MRRs. Our mass and radius estimates modestly favor models with 'thick' H envelopes. However, one star (40 Eri B) is best explained by a 'thin' H envelope. Our results are also consistent with degenerate carbon and carbon/oxygen cores. The general good agreement notwithstanding, there is obvious room for improvement. One significant improvement in the quality of the data will involve additional systems of the type analyzed here, but covering a wider range of masses. For example, we have only one star beyond 0.68 $M_\odot$ (Sirius B). Adding as few as half a dozen more massive white dwarf binaries beyond 0.7 $M_\odot$ would greatly improve this situation. Given the nature of the initial-final-mass relation and the strong dependence of main sequence ages on mass, such massive white dwarfs are likely to be found among Sirius-Like systems. Plans are currently underway to identify such systems and acquire the necessary observations. Realizing the potential of white dwarfs in wide binaries such additional systems depend critically on more parallaxes with improved accuracy. Some of these parallaxes can be expected from on-going ground-based programs such RECONS (Henry et al. 2006) in the southern hemisphere and the US Naval Observatory (USNO) program (Harris et al. 2006) in the northern hemisphere. Others will follow from space-based astrometric missions now in development.

*The JMAPS mission*

The *Joint Milli-Arcsecond Pathfinder Survey* (*JMAPS*) mission is a United States Department of Navy space-based all-sky astrometric survey scheduled for launch in 2014. Among the data provided by *JMAPS* are trigonometric parallaxes accurate to 1 mas for stars down to $12^{th}$ magnitude (Dorland & Dudik 2009). In the context of this paper *JMAPS* parallaxes represent potential improvement over *Hipparcos* parallaxes for the brighter main sequence components of Sirius-like systems prior to the availability of *GAIA* parallaxes, but not a substantial increase in the number or variety of white dwarfs available to test the MRR.

*The GAIA mission*

*GAIA* (Garcia-Berro et al. 2005) is an ambitious astrometric mission planned for launch by the *European Space Agency (ESA)* in mid-2013 as a follow-up to the *Hipparcos* mission. It will operate as an all-sky survey producing positions, proper motions, parallaxes and photometry for up to one billion stars. Here we focus on the absolute trigonometric parallaxes of white dwarfs that *GAIA* will provide. For example, anticipated *GAIA* performance will yield parallaxes of 26 µas for blue stars at magnitude 16. In general, *GAIA* will yield white dwarf parallaxes some 20 to 50 times more precise than most existing ground-based and *Hipparcos* parallaxes. The precision of parallaxes will be a cumulative product of the mission, and most are expected to become available in final form some eight years after launch. Effectively *GAIA* will remove parallax distances as a significant error in radius determinations for white dwarfs out to a distance of 100 pc. Radii uncertainties will then be dominated by photometry. Moreover, during the lifetime of the *GAIA* mission orbital motions exceeding 5 µas yr$^{-1}$ are potentially detectible and could be used to refine orbital motion in wide binaries.

An important consideration regarding the stars 40 Eri B, ε Ret B, Sirius B and CD-38º10980 in Fig. 1 is that the mass-radii error ellipses have little dependence on parallax uncertainties since the relative parallax uncertainties for each star are less than 1%. Indeed, setting the parallax uncertainties to zero has little effect on their error ellipses; other non-parallax uncertainties dominate. Effectively, the parameters of these stars represent the state of affairs that can be expected when *GAIA* parallaxes are available. This fortunate situation is what will potentially exist for a large number of similar systems in the post-*GAIA* era, if the appropriate observations are collected now.

*Current and Future Prospects*

Before new and improved parallax observations become available what are the prospects for additional observations to improve the testing of the MRRs? We are aware of several accepted Cycle 19 *HST* programs aimed at improving the observational status of Sirius B and five additional Sirius-like systems. These observations could well add several new data points to the analysis discussed here. It is also possible that existing ground-based parallax programs may provide new or improved parallaxes that could result in adding new systems or improving the analysis of existing systems. However, truly expanding the number of systems will require parallaxes of the quality promised by *GAIA* and to make optimal use of such parallaxes a focused program of ground-based observing is necessary to provide the supplementary photometry, Balmer line spectroscopy and gravitational redshifts.

A substantial number of such candidate systems exist. These systems presently lack adequate parallaxes, good photometry, spectroscopy or gravitational redshifts. For example the average V-band uncertainty in Table 1 is about 4%. There are several ways to reduce this uncertainty. First is to employ accurate multi-band photometry to compute a solid angle (see Eqn. 4). For DA stars Holberg & Bergeron (2006) have shown that various optical and near IR photometric systems can be placed on a single self–consistent absolute photometric scale that, on average approaches, the 1% level. Going beyond this will require an improved fundamental definition of absolute stellar fluxes, particularly in the near IR. Programs to establish improved absolute flux standards are underway and may produce results in the next few years. A second way to improve the photometry is the availability of high quality near all-sky photometric surveys. Photometric programs such as Pan-STARRS in the northern hemisphere and the VISTA surveys in the southern hemisphere can help provide accurate multi-band photometry for many of these systems over much of the sky. Ultimately such programs as LSST will have design objectives of routinely obtaining 1% photometry. These efforts largely eliminate the need for dedicated photometric observing programs aimed at specific white dwarfs in binary systems.

For the spectroscopic analysis of HI Balmer lines, resolutions of 6 Å or better and signal-to-noise of 60 will suffice to produce good $T_{eff}$ and log g results. For many DA stars good quality spectra already exist but for those having low quality or no spectra, 2m to 4m telescopes are adequate. However, for many systems gravitational redshifts will require 8m telescopes with spectral resolutions of 10,000 or better, and are by far the largest part of any effort. Such programs to both re-observe gravitational redshifts and observe new ones, at the 1 km s$^{-1}$ level are now in the planning stages for both hemispheres. In summary, with a modest amount of observational effort over the next four or five years most of these systems can be observed and vetted as MRR test candidates. It should then be possible to thoroughly test the MRR relations over nearly the full range of white dwarf masses, probing the ratio of stars with 'thick' or 'thin' H envelopes or looking for evidence of stars having exotic core compositions.

Acknowledgements: J.B.H. acknowledges support from NSF grant AST-1008845. T.D.O. acknowledges support from NSF grant AST-087919. Initial portions of this work by J.B.H were also supported by a SIM Science Studies grant (JPL contract no. 1290779). M.A.B. acknowledges the support of the Science and Technology Facilities Council, UK. This research has made use of the *White Dwarf Catalog* maintained at Villanova University and the *SIMBAD* database, operated at CDS, Strasbourg, France.

# REFERENCES

Adams, W. S. 1925, Proc. Natl. Acad. Sci., 11, 382

Barstow, M. A., Bond, H. E., Holberg, J. B., Burleigh, M. R., Hubeny, I., & Koester, D. 2005, MNRAS, 362, 1134

Benvenuto, O. G., Cóisco, A. H., Althouse, L. G., & Serenelli, A. M. 2002, MNRAS, 332, 399

Bergeron, P., Saffer, R. A., & Liebert, J. 1992, ApJ, 394, 228

Bergeron, P., Wesemael, F., & Beauchamp, A. 1995, PASP, 107, 1047

Bergeron, P., Ruiz, M. T., & Leggett, S. K. 1997, ApJS, 108, 339

Bergeron, P., Leggett, S. K., & Ruiz, M. T. 2001, ApJS, 133, 413

Bohlin, R. C., & Gilliland, R. L. 2004, AJ, 127, 3508

Bond, H. E. 2009, Journal of Physics: Conference Series, 172, 012029

Bragaglia, A., Renzini, A., & Bergeron, P. 1995, ApJ, 443, 735

Butler, R. P., Tinney, C. G., Marcy, G. W., Jones, H. R. A., Penny A. J., & Apps K. 2001, ApJ, 555. 410

Butler, R. P., et al. 2006, ApJ, 646, 505

Chauvin, G., Lagrange, A. -M., Udry, S., Fusco, T., Galland F., Naef, D., Beuzit, J. -L., Mayor, M. 2006, A&A, 456, 1165

Dorland, B., & Dudik, R. 2009, arXiv:0907.5248v1

Falcon, R. E., Winget, D. E., Montgomery, M. H., & Williams, K. A. 2010, ApJ, 712, 585

Farihi, J., Burleigh, M. R., Holberg, J. B., Casewell, S. L., & Barstow, M.A. 2011, MNRAS, 417, 1735

Fontaine, G., Brassard, P., & Bergeron, P. 2001, PASP, 113, 409

Garcia-Berro, E., Torres, S., Figueras, F., & Isern, J. 2005, in *White Dwarfs: Cosmological and Galactic Probes*, E. M. Sion, S. Vennes and H. L. Shipman (eds.) Springer, pp15-24

Gianninas, A., Bergeron, P., & Fontaine, G. 2005, ApJ, 631, 1100

Greenstein, J. L., & Trimble, V. L. 1967, 149, 283

Hamada, T. & Salpeter, E. E. 1961 ApJ, 134, 683

Harris, H. C., et al. 2007, AJ, 133, 631

Heintz, W. D. 1974, AJ, 79, 819

Henry, T. J., Jao, W.-C., Breaulieu, T. D., Ianna, PA. Coata E., Méndez, R. A., Subasavage, J. P. 2006, AJ, 132, 2360

Holberg, J. B., & Bergeron, P. 2006, AJ, 132, 1221
Holberg, J. B., Sion, E. M., Oswalt, T., McCook, G. P., Foran, S., & Subsavage, J. P. 2008, AJ, 135, 1225
Holberg, J. B. 2009, *Journal for the History of Astronomy*, 40, 137
Holberg, J. B. 2010, *Journal for the History of Astronomy*, 41, 41
Koester, D. 1987, ApJ, 322, 852
Liebert, J., Bergeron, P., & Holberg, J. B. 2005, ApJS, 156, 47
Landolt, A. U. 1992, AJ, 104, 340
Maxted, P.F.L., O'Donoghue, D., Morales-Rueda, L., Napiwotzki, R., & Smalley, B. 2007, MNRAS, 376, 919
McCook, G. P., & Sion, E. M. 1999, ApJS, 121, 1
Nauenberg, M. 2008, *Journal for the History of Astronomy*, 2010, 39, 297
Oswalt, T. D., Hintzen, P. M., Luyten, W. J. 1988, ApJS, 66, 391
Parsons, S. G., Marsh, T. R., Copperwheat, C. M., Dhillon, V. S., Littlefair, S. P., Gänsicke, B., & Hickman, R. 2010, MNRAS, 402, 2591
Pech, D., Vauclair, G., & Dolez, N. 2006, A&A, 446, 223
Perryman, M. A. C. 1997, The *Hipparcos* and *Tycho* Catalogues (ESA SP-1200; Noordwijk: ESA)
Provencal, J. L., Shipman, H. L., Høg, E. & Thjell, P. 1998 ApJ, 494, 759
Provencal, J. L., Shipman, H. L., Koester, D., Wesemael, F., Bergeron, P. 2002, ApJ, 568, 324
Reid, I. N. 1996, AJ, 111, 2000
Schaefer, G. H., Bond. H. E., Barstow, M., Burleigh, M., Gilliland, R. L., Girard, T. M., Gudehus, D. H., Holberg, J.B., & Nelan E. 2006, BAAS, 38, 1104
Schmidt, H. 1996, A&A, 311, 852
Schulz, H. 1977, A&A, 54, L315
Shipman, H. L., Provencal, J. L., Høg, E., & Thejll, P. 1997, ApJ, 488, L43
Silvestri, N. M., Oswalt, T. D. & Hawley, S. L., Wood, M. A., Smith, J. A., Reid, I. N., & Sion, E. M. 2001, AJ, 121, 503
Smart, R. L. et al. 2003, A&A, 404, 317
Tremblay, P.-E., Ludwig, H.-G., Steffen, M., Bergeron, P., & Freytag, B. 2011, A&A, 531, L19
Trimble, V. & Greenstein, J. L. 1972, ApJ, 177, 441
Van Altena, W. F., Lee, T. J., & Hoffleit, E. D. 1994, General Catalog of Trigonometric Stellar Parallaxes (4$^{th}$ ed. New Haven, CT: Yale Univ. Obs.)
van Leeuwen, F. 2007, *Hipparcos, the New Reduction of the Raw Data,* Springer

Vauclair, G., Schmidt, H., Koester, D., & Allard, N. 1997, A&A, 325, 1055
Wegner, G., & Reid, I. N. 1991, ApJ, 375, 674
Wesemael, F. 2009, Ann. of Sci., 67, 205
Winget D. E., & Kepler, S. 2008, ARA&A, 46, 157
Wood, M. A. 1995, in Proc. 9$^{th}$ European Workshop on White Dwarfs ed. D. Koester & K. Werner (Berlin: Springer)

**Table 1**
**Selected Binary Systems – White Dwarfs**

| WD | Alt ID | Sp Type | $T_{eff}$ (K) | σ | log g | σ | ref | V | σ | ref | π (mas) | σ | ref | $V_{gr}$ (km s$^{-1}$) | σ | ref |
|---|---|---|---|---|---|---|---|---|---|---|---|---|---|---|---|---|
| WD0413-077 | 40 Eri B | DA3.1 | 16402 | 90 | 7.85 | 0.028 | 1 | 9.527 | 0.02 | 7 | 200.65 | 0.23 | 11,12 | 25.8 | 1.4 | 13 |
| WD0416-594 | ε Ret B | DA3.3 | 15310 | 350 | 7.98 | 0.02 | 2 | 12.5 | 0.05 | 2 | 54.84 | 0.5 | 12 | 30.3 | 1.9 | 2 |
| WD0642-166 | Sirius B | DA2 | 25193 | 37 | 8.556 | 0.01 | 3 | 8.44 | 0.06 | 3 | 380.11 | 1.26 | 11,12 | 80.42 | 4.83 | 3 |
| WD1105-048 | LP 672-1 | DA3.3 | 15141 | 88 | 7.848 | 0.018 | 4 | 13.071 | 0.002 | 8 | 22.5 | 7.2 | 11 | 20.1 | 3.2 | 14 |
| WD1143+321 | G148-7 | DA3.4 | 14938 | 96 | 7.929 | 0.018 | 4 | 13.646 | 0.027 | 10 | 31.6 | 2.3 | 11 | 30.8 | 1.2 | 13 |
| WD1314+293 | HZ 43 | DA1.0 | 49000 | 2000 | 7.7 | 0.2 | 4 | 12.914 | 0.01 | 9 | 15.3 | 2.9 | 11 | 30.1 | 3 | 13 |
| WD1327-083 | W485 A | DA3.6 | 13920 | 167 | 7.86 | 0.038 | 5 | 12.327 | 0.019 | 9 | 60.3 | 2.3 | 11,12 | 24.9 | 3.2 | 14 |
| WD1620-391 | CD-38° 10980 | DA2.1 | 24406 | 328 | 8.099 | 0.038 | 6 | 11.008 | 0.002 | 9 | 78.21 | 0.37 | 12 | 33.9 | 0.4 | 15 |
| WD1706+332 | G 181-B5B | DA3.9 | 12960 | 156 | 7.80 | 0.038 | 5 | 15.92 | 0.02 | 10 | 14.85 | 0.807 | 12 | 29.0 | 0.8 | 13 |
| WD1716+020 | LHS 3278 | DA3.8 | 13210 | 159 | 7.77 | 0.038 | 5 | 14.365 | 0.02 | 9 | 28.1 | 2.6 | 11 | 28.3 | 0.9 | 13 |
| WD1743-132 | G154-B5B | DA4.1 | 12300 | 148 | 7.88 | 0.038 | 5 | 14.22 | 0.02 | 10 | 29.96 | 3.86 | 12 | 22.6 | 2.1 | 13 |
| WD2341+322 | LP 347-6 | DA4.0 | 12570 | 151 | 7.93 | 0.038 | 5 | 12.932 | 0.051 | 9 | 56.8 | 1.8 | 11,12 | 29.8 | 1.3 | 13 |

**References.** (1) Bergeron, Saffer & Liebert (1992); (2) Farihi et al. (2011); (3) Barstow et al. (2005); (4) Liebert el al. (2005); (5) Gianninas et al. (2005); (6) Bragaglia et al. (1995); (7) Oswalt et al. (1996); (8) Landolt et al. (1992); (9) Holberg & Bergeron (2006); (10) Weighted Mean; (11) Van Altena et al. (1994); (12) van Leeuwen (2007); (13) Reid (1996); (14) Koester (1987); (15) Silvestri et al. (2001)

**Table 2 Selected Binary Systems – Companions**

| WD (system) | Comp. | Sp Type | V | Sep. (") | a(AU) | Period (yrs) |
|---|---|---|---|---|---|---|
| WD0413-077 | 40 Eri C | M5Ve | 11.17 | 6.943[a] | 34.6[a] | 252.1[a] |
| WD0416-594 | ε Ret | K2IV | 4.442 | 13 | 237 | 3158 |
| WD0642-166 | Sirius A | A1V | -1.44 | 7.5[a] | 19.73[a] | 50.0[a] |
| WD1105-048 | LP672-2 | M3V | 12.55 | 279 | 12410 | $1.6 \times 10^6$ |
| WD1143+321 | G148-6 | M3V | 11.04 | 9.05 | 287 | 4911 |
| WD1314+293 | HZ43 B | M3Ve | 12.66 | 2.2 | 144 | 1670 |
| WD1327-083 | Wolf 485 B | M4V | 14.18 | 501.9 | 8323 | $8.9 \times 10^5$ |
| WD1620-391 | CD-38° 10983 | G5V | 5.376 | 345 | 4412 | $2.3 \times 10^5$ |
| WD1706+332 | BD-33° 2834 | F8V | 9.7 | 35.6 | 465 | 7602 |
| WD1716+020 | LHS 1379 | M3.5V | 12.95 | 13.1 | 684 | $1.9 \times 10^4$ |
| WD1743-132 | G154-B5A | M1V | 11.91 | 32.2 | 1074 | $3.6 \times 10^4$ |
| WD2341+322 | LP347-5 | M1.5V | 12.932 | 173.8 | 3060 | $1.6 \times 10^5$ |

[a]Angular and physical separations and orbital periods from astrometric orbits.

**Table 3 Mass and Radius Summary**

| WD (system) | Mass | σ | Radius | σ | $R_\pi$ | σ | $M_\pi$ | $M_{gr}$ | $M_{red}$ | $\chi^2$ |
|---|---|---|---|---|---|---|---|---|---|---|
| WD0413-077 | 0.51 | 0.036 | 0.0135 | 0.0008 | 0.0135 | 0.00014 | 0.58 | 0.53 | 0.56 | 2.92 |
| WD0416-594 | 0.62 | 0.056 | 0.0133 | 0.0006 | 0.0132 | 0.00040 | 0.60 | 0.60 | 0.62 | 0.18 |
| WD0642-166 | 0.94 | 0.05 | 0.0084 | 0.0025 | 0.00827 | 0.00019 | 1.00 | 0.97 | 1.02 | 4.18 |
| WD1105-048 | 0.45 | 0.094 | 0.0133 | 0.0026 | 0.02502 | 0.00801 | 0.26 | 0.53 | 0.48 | 2.34 |
| WD1143+321 | 0.71 | 0.072 | 0.0149 | 0.001 | 0.0138 | 0.00102 | 0.56 | 0.57 | 0.62 | 2.05 |
| WD1314+293 | 0.80 | 0.25 | 0.0171 | 0.0047 | 0.0163 | 0.00158 | 0.60 | 0.56 | 0.67 | 0.98 |
| WD1327-083 | 0.53 | 0.079 | 0.0141 | 0.00085 | 0.0141 | 0.00057 | 0.54 | 0.53 | 0.55 | 0.11 |
| WD1620-391 | 0.68 | 0.016 | 0.0127 | 0.0028 | 0.0128 | 0.00014 | 0.66 | 0.69 | 0.67 | 1.21 |
| WD1706+332 | 0.54 | 0.085 | 0.0125 | 0.001 | 0.01154 | 0.00064 | 0.71 | 0.50 | 0.60 | 24.6 |
| WD1716+020 | 0.65 | 0.08 | 0.0151 | 0.0015 | 0.01231 | 0.00115 | 0.65 | 0.48 | 0.59 | 19.4 |
| WD1743-132 | 0.46 | 0.11 | 0.0129 | 0.0018 | 0.0130 | 0.00158 | 0.60 | 0.54 | 0.51 | 0.03 |
| WD2341+322 | 0.56 | 0.053 | 0.0124 | 0.0007 | 0.01215 | 0.00049 | 0.66 | 0.57 | 0.61 | 3.9 |

Fig. 1 Observational constraints on the MRRs for the white dwarfs in our sample. In each plot, 1-σ and 2-σ mass-radius contours are compared with theoretical evolved MRRs to the observed $T_{eff}$. For WD0413-077 (1a) and WD0642-166 (1c) we show the range of dynamical masses of Shipman et al. (1997) and Schaefer et al. (2006), respectively. For WD0413-077 a 'thin envelope' MRR model (dotted line) better matches the observations than the 'thick envelope' MRR relation.

Fig. 2 The relative observational constraints on the MRRs for all twelve stars in Fig. 1. The mass-radius plane is defined so that the theoretical radii are set to unity for each star, with the masses unchanged. On this relative radius scale the best fitting radius is divided by the theoretical radius and is plotted as a ratio. Likewise the corresponding 1-σ error ellipses for each star from Fig. 1 are characterized by two dimensional error bars that match the extent and orientation of the original 1-σ error ellipses in terms of relative radii. Error bars are not plotted for three stars, WD1105-048, WD1314+293and WD1743-132, since their uncertainties are so large.

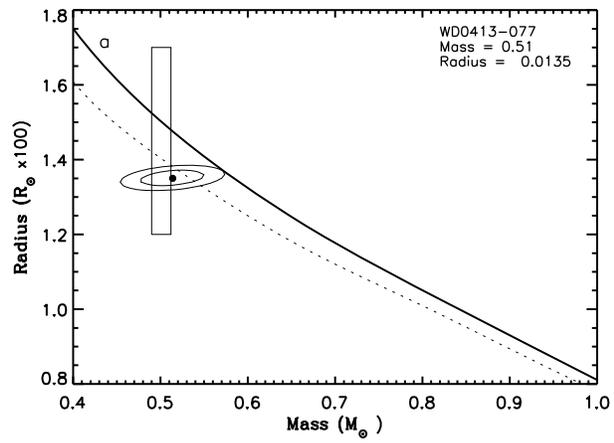
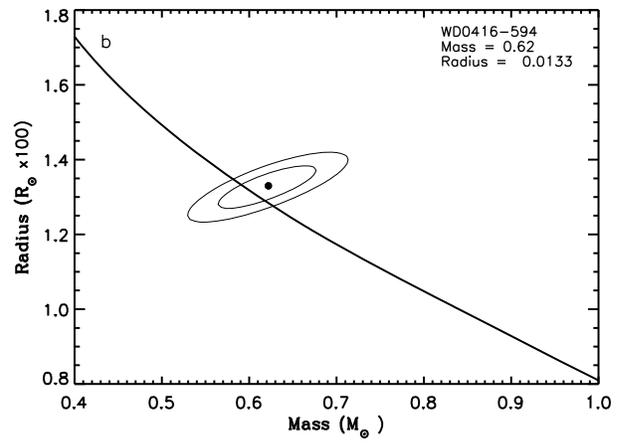
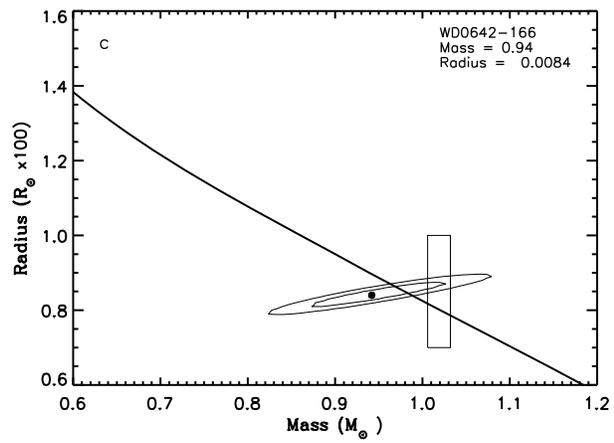
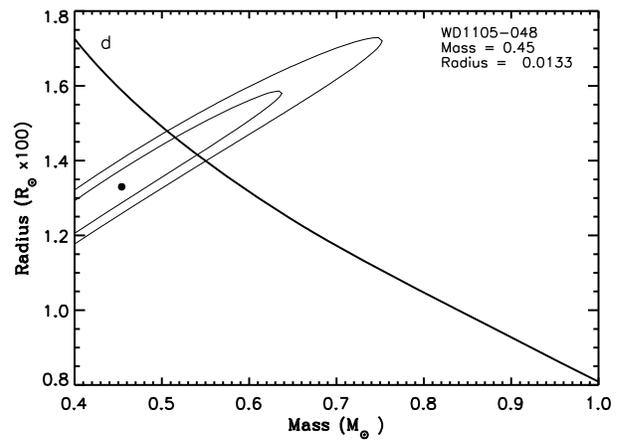

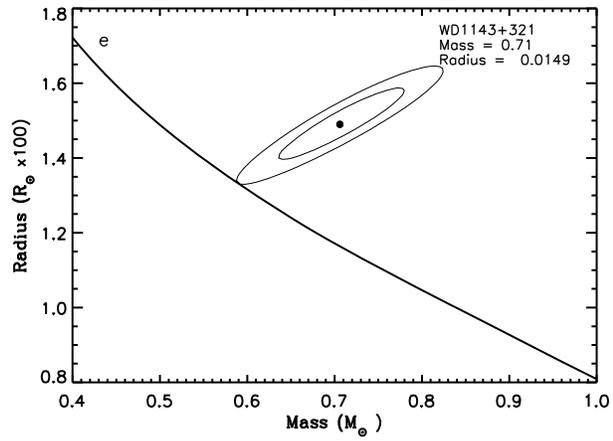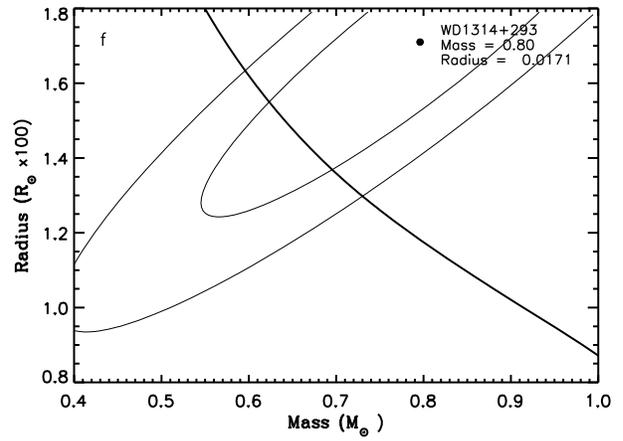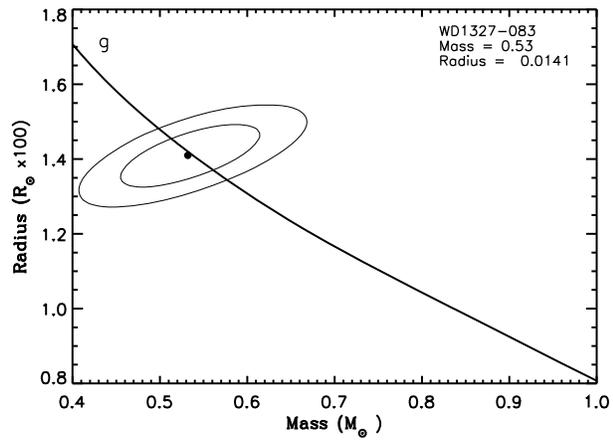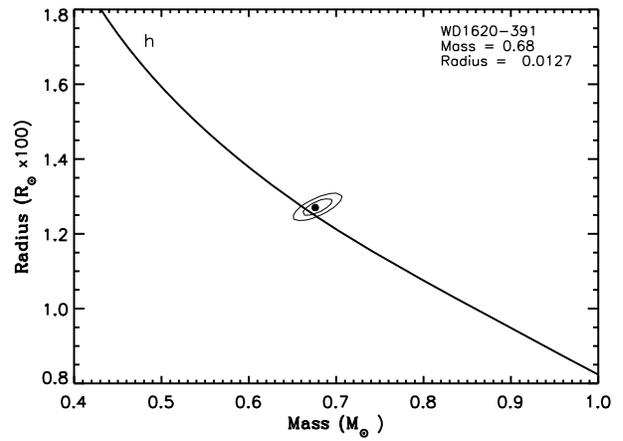

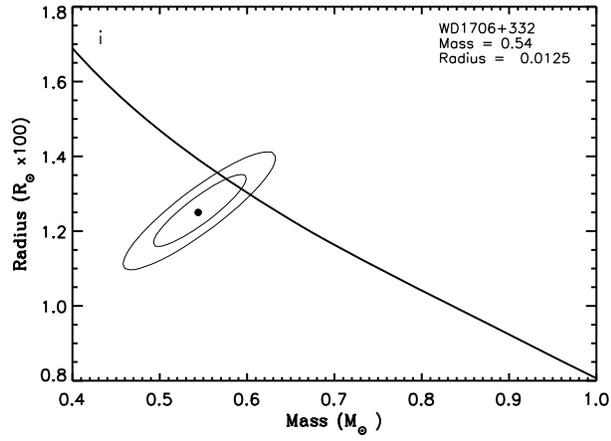
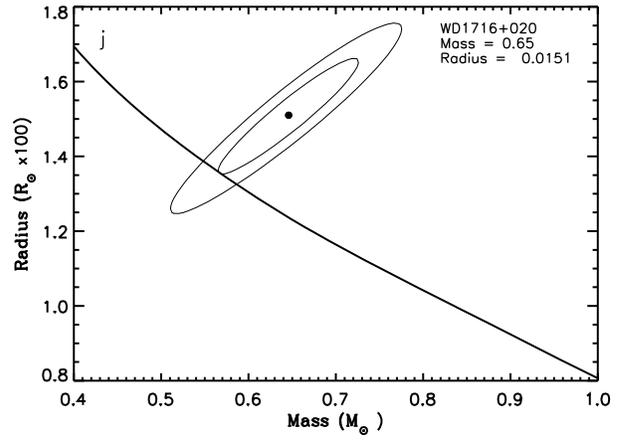
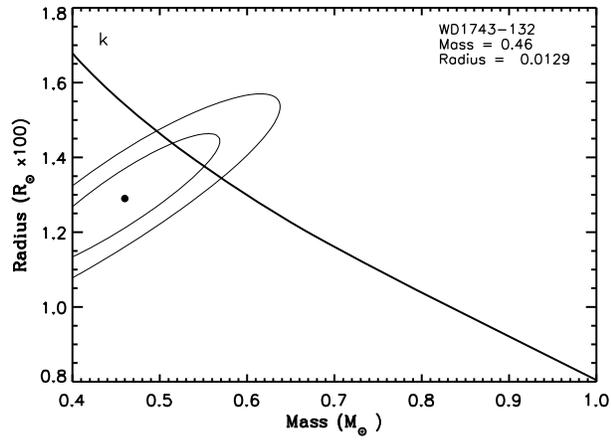
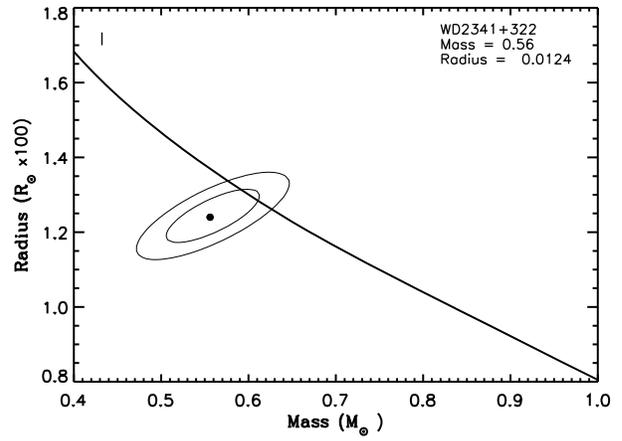

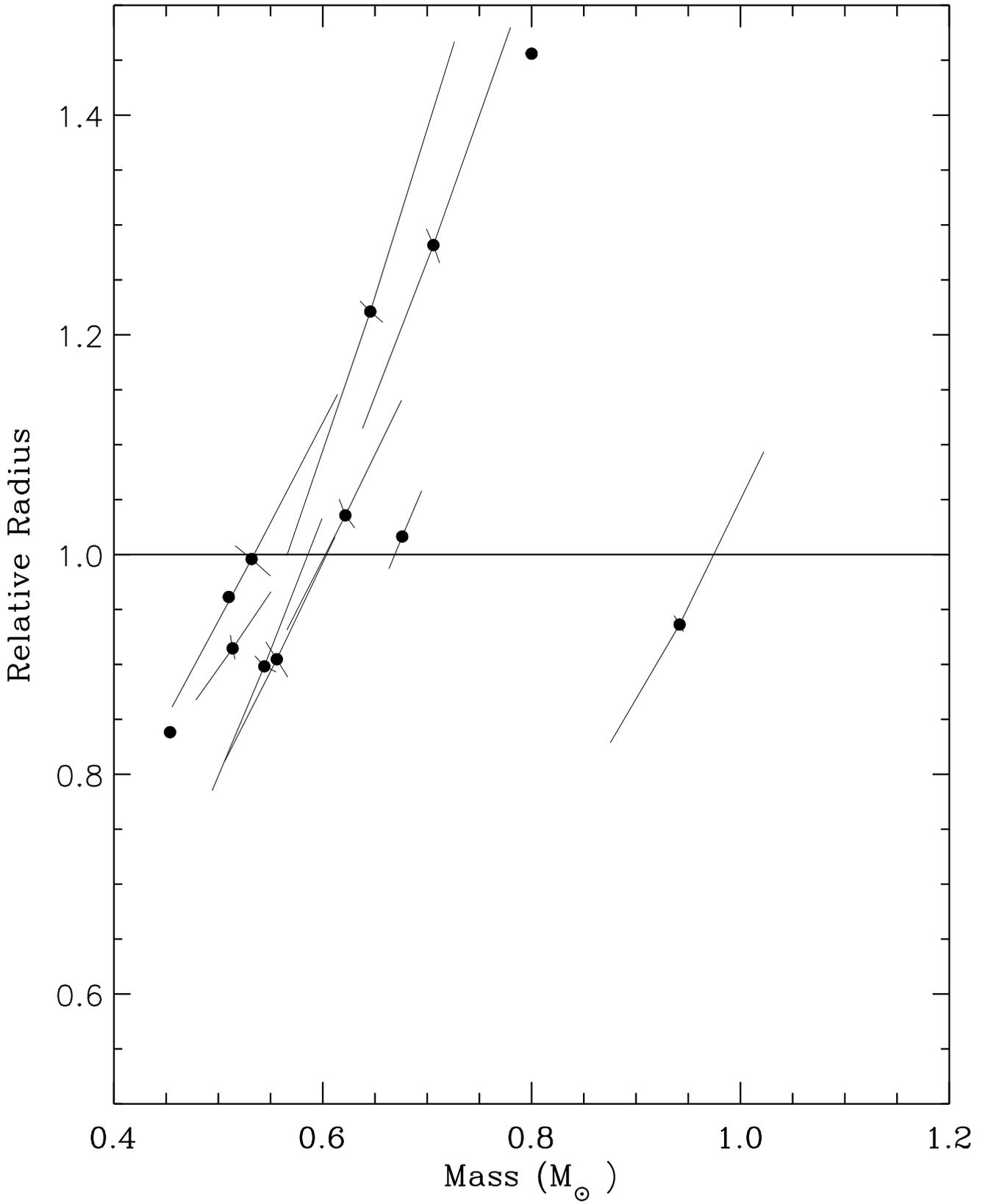